\documentclass[a4paper,11pt]{article}
\usepackage{jheppub}
\makeatletter
\gdef\@fpheader{}
\makeatother
\usepackage{bm,booktabs}
\graphicspath{{./}}

\newcommand{\sthree}{\sigma^{(3)}}
\newcommand{\sT}{\sigma^{(3)}}
\newcommand{\rh}{r_{\mathrm h}}
\newcommand{\qr}{\widehat{\rho}}

\title{Nonlinear pole topology of slow currents in holographic probe-brane metals}

\author[a]{Yan Han}
\author[b,1]{and Xiang-Qian Li\note[1]{Corresponding author.}}
\affiliation[a]{General Education College, Shanxi Institute of Science and Technology,\\
Jincheng 048000, China}
\affiliation[b]{College of Physics and Optoelectronic Engineering, Taiyuan University of Technology,\\
No. 79 West Street Yingze, Taiyuan 030024, China}
\emailAdd{hanyan@sxist.edu.cn}
\emailAdd{lixiangqian@tyut.edu.cn}

\abstract{Low-frequency Drude response does not by itself determine which nearly conserved sector carries an electric current. We classify the leading cubic retarded response of an isolated, inversion-symmetric slow vector by its pole topology. Constitutive nonlinearity produces only incoming-frequency poles, whereas nonlinear relaxation adds a propagator at the emitted frequency; coupling to a second scalar slow mode generates pair-frequency poles. We then compute the full complex third-harmonic response of finite-density Dirac--Born--Infeld probe branes in Lifshitz black-brane backgrounds. In the marginal $z=2$ theory, the dense infrared response approaches the relaxation topology, with a signed infrared weight $p_{\rm IR}=0.994\pm0.002$ and a normalization-independent zero at $\omega_\times/\Gamma_J=0.2918$. Nonmarginal $z=1$ and $z=3/2$ backgrounds instead realize mixed constitutive--relaxation weights, demonstrating that the topology is selected dynamically rather than imposed by the DBI square root. The exact nonlinear dc solution further yields a parameter-free dense-limit scaling function and a crossover field $E_{\rm nl}\propto T^{3/2}$ at $z=2$. These results connect emergent higher-form slow currents, nonlinear response theory, and holographic transport, while the two-mode extension identifies the additional pole structures expected when energy or deformation modes remain slow.}


\begin{document}
\maketitle
\flushbottom

\section{Introduction}

Holographic transport supplies controlled access to real-time response in strongly coupled systems without quasiparticles. Finite-density probe branes are especially useful because their electric sector can develop a parametrically narrow Drude peak even when momentum does not participate in the transport channel \cite{Karch2007,KarchOBannonThermo2007,HartnollPolchinski2010,Chen2017,Gushterov2018}. Nonlinear dc transport in related holographic systems has been characterized through open-string horizons and fully backreacted steady states \cite{Karch2011,KimPang2011,Horowitz2013,Withers2016,Ishigaki2024}. The associated slow current has been interpreted in terms of an approximate higher-form symmetry and its mixed anomaly with the ordinary electric charge \cite{Gaiotto2015,Grozdanov2017,IqbalMacfarlane2021,Davison2023}. This symmetry-based description accounts for the long lifetime of the current, but its consequences for nonlinear frequency-dependent response have not been established.

The missing information is invisible in a linear Drude form. Momentum, quasiparticles, phase-relaxed superfluid sectors, fluctuating density waves, and emergent higher-form currents can all produce a response characterized by one weight and one relaxation time \cite{Hartnoll2015,LucasSachdev2015,Davison2016,Delacretaz2017PRB,LucasFong2018,HartnollMackenzie2022}. At third order, however, the analytic structure also records where the nonlinearity enters the slow dynamics. Nonlinear constitutive relations dress the incoming fields, nonlinear relaxation propagates the emitted harmonic, and feedback through another slow sector introduces poles at pairwise sums of the incoming frequencies. We refer to this pattern of slow denominators as the \emph{pole topology}. Related outgoing propagators occur in nonlinear electron hydrodynamics \cite{Sun2018}, while multidimensional spectroscopy provides a natural way to resolve pair-frequency structures \cite{Wan2019,Parameswaran2020,Choi2020,Barbalas2025,Liu2025}.

In this paper we first derive the leading cubic response of a single approximately conserved vector and its minimal current--energy extension. We then solve the finite-density DBI fluctuation problem to cubic order in a Lifshitz black-brane background. The marginal $z=2$ model is analytically favorable: its linear relaxation time, nonlinear dc conductivity, and dense-limit scaling function can all be obtained exactly. The full complex numerical response shows that this background approaches a nearly pure relaxation topology at high density. By repeating the calculation at $z=1$ and $z=3/2$, we show that other members of the same DBI family realize genuine mixtures of constitutive and relaxation nonlinearities. This comparison is important: the DBI square root permits both topologies, while the infrared geometry and state select their relative weight.

The resulting framework also interfaces with strange-metal phenomenology without using material data as an input to the holographic calculation. A slow energy or deformation mode produces pair-frequency poles of the type found in kinetic descriptions of Yukawa--Sachdev--Ye--Kitaev metals \cite{Esterlis2021,Guo2022,Patel2023,Li2024,Kryhin2025}. As an illustration, independently reported current and energy relaxation rates in normal-state LSCO select a frequency window for the minimal two-mode response \cite{Chaudhuri2026}; the broader normal-state transport context is documented in Refs.~\cite{Cooper2009,Legros2019}. We keep this application explicitly separate from the holographic result: it tests the pole-topology diagnostic, not a microscopic identification of LSCO with the probe-brane theory.

The paper is organized as follows. Section~\ref{sec:eft} derives the slow-mode effective theory, including its current--energy and general multimode extensions. Sections~\ref{sec:dbi} and \ref{sec:scaling} develop the fixed-density holographic calculation, establish its frequency-dependent topology, and obtain the exact nonlinear scaling limit. Section~\ref{sec:phenomenology} discusses the two-mode experimental diagnostic, and section~\ref{sec:discussion} summarizes the scope and limitations.

\section{Slow-current effective theory and pole topology}
\label{sec:eft}

\subsection{Single-current response}
Let $u_i$ be the only homogeneous vector whose relaxation rate
$\Gamma_J$ is parametrically smaller than a microscopic rate
$\Lambda_{\rm fast}$.
For an isotropic inversion-symmetric state, the intrinsic slow-sector equations through cubic order are
\begin{align}
 j_i&=\chi u_i+\alpha (u_ku_k)u_i
      +O(u^5,\partial_tu),\label{eq:eftj}\\
 (\partial_t+\Gamma_J)u_i+\beta(u_ku_k)u_i
 &=\kappa E_i+O(u^5,\partial_t^2u).\label{eq:eftu}
\end{align}
The coefficients are real in the time domain but need not have either sign.
Noise and Schwinger--Keldysh partners are required for fluctuation observables, but do not alter the retarded tree-level pole counting used here.

For an emergent higher-form current, approximate higher-form charge conservation provides the slow variable and the mixed anomaly with ordinary charge gives the linear electric driving.
Neither symmetry fixes $\alpha$ or $\beta$.
Moreover, the most general effective theory can contain source nonlinearities such as
$E^2E_i$, $(u\cdot E)E_i$, and $u^2E_i$.
Such terms generate contact contributions or terms with fewer slow propagators.
Consequently, Eqs.~\eqref{eq:eftj}--\eqref{eq:eftu} classify the \emph{leading pole-enhanced} cubic response in the limit
\begin{equation}
 \Gamma_J,\ |\omega_a|\ll\Lambda_{\rm fast},
 \qquad \omega_a/\Gamma_J\ {\rm fixed}.
\label{eq:scalinglimit}
\end{equation}
Accordingly, the classification refers to the pole-enhanced scaling limit; the response at microscopic frequencies also contains nonuniversal contact and derivative terms.

Write
$E(t)=\sum_a E_a e^{-i\omega_at}$,
$D_\omega=\Gamma_J-i\omega$, and
$\Omega=\omega_1+\omega_2+\omega_3$.
The linear solution is
\begin{equation}
 u^{(1)}(\omega_a)=\frac{\kappa E_a}{D_{\omega_a}}.
\end{equation}
At the sum frequency, the cubic correction generated by nonlinear relaxation is
\begin{equation}
 u^{(3)}(\Omega)=
-\frac{\beta\kappa^3}
 {D_\Omega D_{\omega_1}D_{\omega_2}D_{\omega_3}}
 E_1E_2E_3 ,
\label{eq:u3}
\end{equation}
up to the conventional permutation factor used to define the symmetric third-order conductivity.
The nonlinear constitutive term gives instead
\begin{equation}
 j_{\alpha}^{(3)}(\Omega)=
\frac{\alpha\kappa^3}
 {D_{\omega_1}D_{\omega_2}D_{\omega_3}}
E_1E_2E_3 .
\label{eq:ja3}
\end{equation}
Combining Eqs.~\eqref{eq:u3} and \eqref{eq:ja3},
\begin{equation}
 \sigma^{(3)}(\Omega;\omega_1,\omega_2,\omega_3)
 =\frac{A}{\prod_aD_{\omega_a}}
 +\frac{B}{D_\Omega\prod_aD_{\omega_a}},
\label{eq:mixed}
\end{equation}
where $A\propto\alpha\kappa^3$ and
$B\propto-\chi\beta\kappa^3$.
After normalization by the dc value,
\begin{equation}
 \frac{\sigma^{(3)}}{\sigma_{\rm dc}^{(3)}}
 =(1-p)\frac{\Gamma_J^3}{\prod_aD_{\omega_a}}
 +p\frac{\Gamma_J^4}{D_\Omega\prod_aD_{\omega_a}},
\quad
p=\frac{B/\Gamma_J^4}{A/\Gamma_J^3+B/\Gamma_J^4}.
\label{eq:pform}
\end{equation}
The quantity $p$ is a signed topology weight normalized by the net dc cubic response, not a probability or a positive spectral fraction.
Opposite signs of $A$ and $B$ can place it outside $[0,1]$.
In particular, $p>1$ means that the relaxation term has the sign of the net response while the constitutive term has weight $1-p<0$ and partially cancels it; $p<0$ describes the converse cancellation.
This possibility is natural because neither symmetry nor entropy production fixes the signs of the static cubic constitutive coefficient $\alpha$ and the nonlinear relaxation coefficient $\beta$ separately.

For third-harmonic generation (THG), $\omega_a=\omega$.
The real part of the constitutive topology vanishes first at
$\omega/\Gamma_J=1/\sqrt3$, whereas the relaxation topology obeys
\begin{align}
 1-12x^2+3x^4&=0,
\nonumber\\
 x_\times&=\sqrt{2-\sqrt{33}/3}
 =0.2917975059\ldots .
\label{eq:zeros}
\end{align}
The low-frequency expansion of Eq.~\eqref{eq:pform} is
\begin{equation}
\frac{\sigma^{(3)}(3\omega;\omega,\omega,\omega)}
 {\sigma_{\rm dc}^{(3)}}
 =1+i\omega\,3(1+p)\tau_{\rm slow}+O(\omega^2),
\qquad \tau_{\rm slow}=\Gamma_J^{-1}.
\label{eq:delayp}
\end{equation}
Thus the infrared topology can be read off without fitting a frequency window:
\begin{equation}
 p_{\rm IR}=\frac{\tau_{\rm THG}}{3\tau_{\rm slow}}-1.
\label{eq:pir}
\end{equation}
This estimator is meaningful when the linear response contains one isolated
infrared pole and all omitted modes remain fast on the frequency window used
to extract the slopes. Agreement of the independently computed phase slopes
with the full complex spectrum, together with convergence in density, tests
whether the single-mode regime is attained.

\subsection{Current--energy and multimode sectors}
The minimal experimentally anchored multimode extension consists of a slow
current amplitude $u$ and an energy-density deviation $\varepsilon$:
\begin{align}
(\partial_t+\Gamma_M)u+\lambda\varepsilon u&=\kappa E,
\label{eq:ceu}\\
(\partial_t+\Gamma_E)\varepsilon&=qEu.
\label{eq:cee}
\end{align}
This is the slow-mode reduction of the familiar electrothermal feedback in
which field-driven current deposits energy and the resulting energy deviation
modifies current relaxation \cite{Allen1987,Barbalas2025}.
The measured current is $j=\chi u$; additional constitutive nonlinearities can
be superposed but are not required for the electrothermal pole structure.
At first order,
\begin{equation}
u^{(1)}(\omega_a)=\frac{\kappa E_a}{D_a},
\qquad D_a=\Gamma_M-i\omega_a.
\end{equation}
At second order the energy response at a pair frequency is
\begin{equation}
\varepsilon^{(2)}(\omega_i+\omega_j)
=\frac{q\kappa E_iE_j}{D^E_{ij}}
\left(\frac{1}{D_i}+\frac{1}{D_j}\right),
\qquad
D^E_{ij}=\Gamma_E-i(\omega_i+\omega_j).
\label{eq:energy2}
\end{equation}
Consequently the symmetric cubic conductivity is proportional to
\begin{equation}
\sigma^{(3)}_{\rm CE}(\Omega;\omega_1,\omega_2,\omega_3)
\propto
\frac{1}{D_\Omega}
\sum_{\rm cyc}
\frac{D_i^{-1}+D_j^{-1}}
 {D^E_{ij}D_k},
\qquad D_\Omega=\Gamma_M-i\Omega.
\label{eq:CEgeneral}
\end{equation}
Equation~\eqref{eq:CEgeneral} displays the pole geometry directly:
three incoming current poles, one emitted-current pole, and three
pair-frequency energy poles.  The latter form distinct ridges in a
mixed-frequency or two-dimensional spectrum and cannot be generated by
renormalizing the single-current weight $p$ \cite{Wan2019,Salvador2024}.

For THG, $\omega_1=\omega_2=\omega_3=\omega$, all three pair channels
coincide.  Normalization by the dc value removes $\lambda$, $q$, $\chi$, and
$\kappa$:
\begin{equation}
\frac{\sigma^{(3)}_{\rm CE}(3\omega)}
 {\sigma^{(3)}_{{\rm CE},\rm dc}}
=\frac{\Gamma_M^3\Gamma_E}
 {(\Gamma_M-3i\omega)(\Gamma_M-i\omega)^2
  (\Gamma_E-2i\omega)}.
\label{eq:CEthgS}
\end{equation}
With $x=\omega/\Gamma_M$ and $g=\Gamma_E/\Gamma_M$, the real-part numerator is
\begin{equation}
6x^4-(7g+10)x^2+g.
\end{equation}
The two positive zeros are therefore
\begin{equation}
x_{\times,\pm}^2=
\frac{7g+10\pm\sqrt{49g^2+116g+100}}{12}.
\label{eq:CEroots}
\end{equation}
The lower zero behaves as $x_{\times,-}\simeq\sqrt{g/10}$ for $g\ll1$,
while the second approaches $\sqrt{5/3}$.
The phase expansion gives
\begin{equation}
\frac{\sigma^{(3)}_{\rm CE}(3\omega)}
 {\sigma^{(3)}_{{\rm CE},\rm dc}}
=1+i\omega\left(\frac{5}{\Gamma_M}
                +\frac{2}{\Gamma_E}\right)+O(\omega^2).
\label{eq:CEphaseS}
\end{equation}

\paragraph{Relation to YSYK energy and deformation channels.}
The pair-frequency topology above is the low-frequency Markovian reduction of
the kinetic structures obtained in the YSYK strange metal
\cite{Kryhin2025}. In that calculation the cubic current is built from a
dipolar propagator at the emitted frequency, an even-parity propagator at a
pair frequency, and dipolar propagators at incoming frequencies. Denoting the
even mode by $X$, the corresponding pole content has the schematic form
\begin{equation}
\sigma_X^{(3)}
\sim \frac{1}{D^M_\Omega}
\sum_{\rm cyc}
\frac{\mathcal T_X(\omega_i,\omega_j,\omega_k)}
{D^X_{ij}D^M_k}
\left(\frac{1}{D^M_i}+\frac{1}{D^M_j}\right),
\qquad
D^X_{ij}=\Gamma_X-i(\omega_i+\omega_j),
\label{eq:ysykmap}
\end{equation}
where $\mathcal T_X$ contains nonsingular amplitudes and the polarization
tensor. Equation~\eqref{eq:CEgeneral} is the scalar, isotropic realization of
Eq.~\eqref{eq:ysykmap}. The correspondence concerns the slow denominators;
the full YSYK kernels retain frequency-dependent scattering and therefore need
not have the parameter-free residues or THG line shape of
Eq.~\eqref{eq:CEthgS}.

For the energy-density channel, $X=E$, the isolated YSYK model has an exact
conservation pole. Coupling to the experimental environment replaces it by
$D^E_{ij}=\Gamma_E-i(\omega_i+\omega_j)$, with $\Gamma_E$ independently
measured by pump--probe relaxation. The finite-$\Gamma_E$ prediction in this
article is therefore an open-system continuation of the heating channel. The
full YSYK response retains frequency-dependent kinetic kernels beyond this
two-rate low-frequency reduction.

For a square lattice the even deformation channels are $X=B_{1g}$ and
$B_{2g}$, with rates $\Gamma_{B_{1g}}$ and $\Gamma_{B_{2g}}$. Their
pair-frequency poles have the same topology as Eq.~\eqref{eq:ysykmap}, but
their tensor residues provide an additional discriminator. Fields along the
principal axes select the $B_{1g}$ response, fields along the diagonals select
$B_{2g}$, and at a generic angle a transverse third-order current contains no
scalar heating contribution. Consequently a practical mode-identification
protocol combines (i) pole locations from phase-resolved frequency sweeps,
(ii) axis-versus-diagonal subtraction, and (iii) the transverse cubic signal.
A THG zero alone cannot determine the deformation rate because different
tensor residues can shift or remove the zero.

\paragraph{General multimode extension.}\label{sec:multimode}
For slow vectors $u_i^A$, linear dynamics takes the matrix form
\begin{equation}
 [(-i\omega)\delta^{AB}+\Gamma^{AB}]u_i^B
 =\kappa^A E_i .
\end{equation}
Every scalar factor $D_\omega^{-1}$ in Eq.~\eqref{eq:mixed} is then replaced by a matrix resolvent
$[(-i\omega)\mathbb 1+\bm{\Gamma}]^{-1}$, contracted with the driving, constitutive, and cubic-relaxation tensors.
The response generally contains several incoming and outgoing poles and cannot be represented by a single $p$.
Deviations from the one-rate form therefore reveal a multimode slow sector, apart from accidental degeneracies.
However, the location of one THG sign zero is not sufficient to reconstruct $\bm{\Gamma}$: distinct rates, residues, and nonlinear tensors can yield the same zero.
Quantitative mode identification requires a fit to both quadratures over frequency, with candidate rates constrained by linear response, and is strengthened by mixed-frequency measurements that vary the incoming and outgoing denominators independently.
In an experimental fit, analytic contact backgrounds should be included separately because they have fewer slow denominators.
Additional narrow structures associated with disorder, phonons, or other nearly conserved operators require enlarging $\bm{\Gamma}$ rather than absorbing them into $p$.

\section{Holographic DBI realization}
\label{sec:dbi}
We now determine which topology is selected by a strongly coupled microscopic theory.
Consider a finite-density DBI gauge field on a Lifshitz black brane, following
the probe-brane strange-metal construction of Ref.~\cite{HartnollPolchinski2010}
and its finite-density DBI antecedents \cite{Karch2007,KarchOBannonThermo2007},
\begin{align}
 S&=-\mathcal N\int d^4x\sqrt{-\det(g+F)},\nonumber\\
 ds^2&=\frac1{r^2}\left[-\frac{fdt^2}{r^{2z-2}}
 +\frac{dr^2}{f}+d\bm{x}^2\right],\nonumber\\
 f(r)&=1-(r/\rh)^{z+2},\qquad
 T=\frac{z+2}{4\pi\rh^z}.
\label{eq:model}
\end{align}
Here $d\bm{x}^2=dx^2+dy^2$.
The probe limit should be understood as an open-sector limit: the charged
flavour degrees of freedom exchange momentum and energy with a parametrically
larger neutral adjoint bath, while the bath's response to the flavour stress
tensor is suppressed.  Consequently, the flavour conductivity can be finite
without explicit relaxation of the total boundary momentum.  The slow pole
studied below belongs to this probe current sector, not to a self-consistent
momentum mode of the full boundary theory.

\subsection{Fixed-density expansion}
For a homogeneous perturbation in the $x$ direction, the nontrivial determinant block in Schwarzschild coordinates is
\begin{equation}
 -\det(g+F)=Gbc+G a_x'^2-b\dot a_x^2-cA_t'^2,
\end{equation}
where $G=|g_{tt}|$, $b=g_{rr}$, and $c=g_{xx}$.
After including the spectator $y$ direction, the DBI Lagrangian is
\begin{equation}
 \mathcal L=-\mathcal N\sqrt c\,
 \sqrt{c(Gb-A_t'^2)+Ga_x'^2-b\dot a_x^2}.
\end{equation}
The conserved density is $\rho=\partial\mathcal L/\partial A_t'$.
Eliminating $A_t'$ by a Legendre transform gives the fixed-density Routhian
\begin{equation}
 \mathcal L_R=-\sqrt{\rho^2+\mathcal N^2c^2}\,
 \sqrt{Gb+\frac Gc a_x'^2-\frac bc\dot a_x^2}.
\label{eq:Routh}
\end{equation}
This order of operations matters: at nonlinear order, holding $A_t'$ fixed is not equivalent to holding the physical charge density fixed.

In units $\mathcal N=\rh=1$, the metric functions are
\begin{equation}
 G=\frac f{r^{2z}},\qquad b=\frac1{r^2f},\qquad c=\frac1{r^2},
\qquad f=1-r^{z+2}.
\end{equation}
Writing $Q=\sqrt{1+\qr^2r^4}$ and expanding Eq.~\eqref{eq:Routh} yields
\begin{align}
 \mathcal L_2&=-\frac P2a_x'^2+\frac W2\dot a_x^2,\nonumber\\
 \mathcal L_4&=\frac R4(Ga_x'^2-b\dot a_x^2)^2,
\end{align}
with
\begin{equation}
 P=fr^{1-z}Q,\qquad W=\frac{r^{z-1}Q}{f},\qquad
 R=\frac12r^{3z+5}Q.
\end{equation}

\subsection{Regular ingoing fluctuation problem}
To impose horizon regularity directly, let
\begin{equation}
 h'(r)=\sqrt{\frac bG}=\frac{r^{z-1}}f,\qquad v=t-h(r)
\end{equation}
be an ingoing coordinate appropriate to a horizon at larger $r$.
For a Fourier component $e^{-i\omega v}\phi(r)$, the Schwarzschild-time radial derivative is
\begin{equation}
 D_\omega\phi=\phi'+i\omega h'\phi.
\end{equation}
The identity $Ph'=Q$ cancels all apparent horizon divergences and turns the linear equation into
\begin{equation}
 (P\phi')'+2i\omega Q\phi'+i\omega Q'\phi=0.
\label{eq:lin}
\end{equation}
At $r=1$, regularity fixes
\begin{equation}
 \phi_h'=-\frac{i\omega Q_h'}{P_h'+2i\omega Q_h}\phi_h.
\end{equation}

For a monochromatic source, the cubic invariant at $3\omega$ can be written
\begin{equation}
 X=G\phi'^2+2i\omega\sqrt{Gb}\,\phi\phi'.
\end{equation}
The source for the third-harmonic profile $\psi$ is
\begin{equation}
\mathcal S_3=
\left[RG(D_\omega\phi)X\right]'
+3i\omega h'RG\phi'X.
\label{eq:source}
\end{equation}
The second term in Eq.~\eqref{eq:source} is the manifestly regular remainder after cancellation between the time derivative and the singular part of $D_\omega\phi$.
The cubic equation is
\begin{equation}
 (P\psi')'+6i\omega Q\psi'+3i\omega Q'\psi=\mathcal S_3.
\label{eq:cubic}
\end{equation}
Its horizon condition is
\begin{equation}
 \psi_h'=\frac{\mathcal S_{3,h}-3i\omega Q_h'\psi_h}
 {P_h'+6i\omega Q_h}.
\end{equation}
We impose $\phi(0)=1$ and $\psi(0)=0$.

The boundary fluxes are
\begin{align}
 j^{(1)}&=\lim_{r\to0}P D_\omega\phi,\nonumber\\
 j^{(3)}&=\lim_{r\to0}\left[
 P D_{3\omega}\psi-RG(D_\omega\phi)X\right],
\end{align}
and the conductivities used in section~\ref{sec:dbi} are
\begin{equation}
 \sigma(\omega)=\frac{j^{(1)}}{i\omega},\qquad
 \sT(3\omega;\omega,\omega,\omega)=\frac{j^{(3)}}{(i\omega)^3}.
\end{equation}

\subsection{Exact dc response and linear relaxation at \texorpdfstring{$z=2$}{z=2}}
For a constant electric field, reality of the DBI solution selects an open-string horizon $r_\ast$ \cite{Karch2007,KimPang2011,Ishigaki2024}:
\begin{equation}
 E^2=\frac{f(r_\ast)}{r_\ast^{2z+2}}.
\end{equation}
The conserved current is
\begin{equation}
 J=E\sqrt{\mathcal N^2+\rho^2r_\ast^4}.
\end{equation}
Writing $r_\ast=\rh[1-\delta+O(E^4)]$ gives
\begin{equation}
 \delta=\frac{E^2\rh^{2z+2}}{z+2}.
\end{equation}
Therefore
\begin{align}
 \sigma_{\rm dc}&=\mathcal N\sqrt{1+\qr^2},\\
 \sT_{\rm dc}&=-\frac{2\mathcal N\qr^2\rh^{2z+2}}
 {(z+2)\sqrt{1+\qr^2}}.
\end{align}
These formulas provide an independent zero-frequency check of the ac calculation.

It is useful to derive the exact low-frequency coefficient without solving the second-order radial equation.
Define the radial conductivity in Schwarzschild coordinates by
\begin{equation}
 \sigma(r,\omega)=\frac{P A_x'}{i\omega A_x}.
\end{equation}
The linear wave equation implies the Riccati flow, in the spirit of the
holographic membrane-paradigm evolution of response functions \cite{IqbalLiu2009},
\begin{equation}
 \sigma'=i\omega\left(W-\frac{\sigma^2}{P}\right),
\qquad \sigma(1,\omega)=Q_h.
\end{equation}
Expanding $\sigma=Q_h+i\omega s+O(\omega^2)$ gives
\begin{equation}
 s(0)=\int_0^1dr\,\frac{Q_h^2-Q(r)^2}{P(r)}.
\end{equation}
For $z=2$, the factor $1-r^4$ cancels:
\begin{equation}
 s(0)=\int_0^1dr\,\frac{\qr^2r}{\sqrt{1+\qr^2r^4}}
=\frac{\qr}{2}\operatorname{arcsinh}(\qr).
\end{equation}
Dividing by $\sigma_{\rm dc}=Q_h$ yields
\begin{equation}
 \tau_J=\frac{\qr\,\operatorname{arcsinh}(\qr)}
 {2\sqrt{1+\qr^2}}.
\end{equation}
Here $\tau_J$ is the dimensionless radial delay in units $\rh=1$;
the physical slow time is $\tau_{\rm slow}=\rh^2\tau_J$ for $z=2$.

\subsection{Numerical implementation and topology selection}
\begin{figure}[t]
\includegraphics[width=\linewidth]{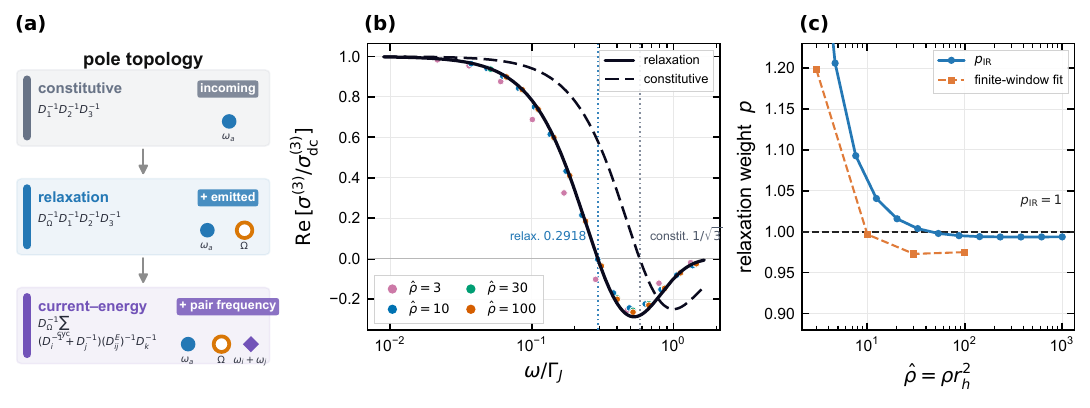}
\caption{\label{fig:topology}
Nonlinear pole topology and its holographic realization.
(a) Schematic slow-propagator factors distinguish constitutive, relaxation, and current--energy topologies; $D_a=\Gamma_J-i\omega_a$, $D_\Omega=\Gamma_J-i\Omega$, and $D^E_{ij}=\Gamma_E-i(\omega_i+\omega_j)$. Each glyph labels a pole family rather than its multiplicity, and the current--energy expression includes the three cyclic choices of $(i,j,k)$.
(b) Normalized dissipative THG of the $d=z=2$ DBI metal; points are full numerical data and the solid (dashed) curve is the parameter-free relaxation (constitutive) EFT limit fixed by $\Gamma_J$.
Dotted lines mark the calibration-independent zeros $\omega_\times/\Gamma_J=0.2918$ and $1/\sqrt3$.
(c) Eighteen two-frequency, Richardson-extrapolated estimates of the signed infrared weight $p_{\rm IR}=\tau_{\rm THG}/(3\tau_J)-1$ (blue) approach the relaxation limit $p_{\rm IR}=1$.
At low density $p_{\rm IR}>1$ indicates a compensating constitutive contribution of opposite sign.
Orange squares are fits over $\omega\leq\Gamma_J$ for the four full frequency scans; their displacement from the blue infrared trend reflects higher-derivative corrections.}
\end{figure}

Equations~\eqref{eq:lin} and \eqref{eq:cubic} are integrated from
$r=1-2\times10^{-5}$ to $r=2\times10^{-5}$ with an eighth-order adaptive Runge--Kutta method.
The cubic source is evaluated on a grid uniform in $-\ln(1-r)$ and represented by a complex cubic spline.
The source-free boundary condition is imposed by combining one homogeneous ingoing solution and one sourced ingoing solution.
For the $z=2$ scans in Fig.~\ref{fig:topology}, each density uses 25 logarithmically spaced frequencies $0.025\leq\omega\leq12$ in units $\rh=1$; the plotted points are then normalized by the exact $\Gamma_J$.
The nonmarginal checks use the corresponding logarithmic grids shown in the archived data files.
The third-order convention is the fully symmetric coefficient defined by
$j^{(3)}(\Omega)=\sthree(\Omega;\omega_1,\omega_2,\omega_3)E_1E_2E_3$ after summing equivalent permutations; no additional $1/3!$ is included in the tabulated THG coefficient.

Three checks were applied:
\begin{enumerate}
\item At $\omega\to0$, both $\sigma$ and $\sT$ converge to the independent exact dc formulas above.
\item Increasing the source grid from 1500 to 6000 points changes the complex $\sT$ at $(\qr,\omega)=(3,0.3)$ by less than $1.3\times10^{-7}$ relatively.
\item The residual cubic boundary source is below $10^{-17}$ throughout the production frequency scans, typically near machine precision.
\end{enumerate}

The low-frequency nonlinear delay was extracted by Richardson extrapolation from two frequencies.
For $\qr\geq30$,
$\tau_{\rm THG}/(6\tau_J)$ differs from unity by less than $0.4\%$.
The parameter-free relaxation spectrum has complex relative errors $0.0378$, $0.0277$, and $0.0218$ for $\qr=10$, $30$, and $100$, while the corresponding numerical sign reversals occur at $\omega_\times/\Gamma_J=0.2910$, $0.2983$, and $0.2979$.
Calculations at $1\le z<2$ also show a regular finite-temperature cubic response, as quantified below.
The $z=2$ marginal model is singled out in this article because it combines an analytic logarithm, $T$-linear resistivity, and a parameter-free linear relaxation rate.

The numerical cubic response distinguishes the two EFT mechanisms decisively.
Over $\omega\leq\Gamma_J$, a pure constitutive spectrum has a complex relative error of $56\%$--$63\%$ for $\qr=10$--$100$, while the relaxation spectrum has only $2\%$--$4\%$ error.
More sharply, Eq.~\eqref{eq:pform} implies
\begin{equation}
 \frac{\tau_{\rm THG}}{\tau_J}=3(1+p_{\rm IR}).
\label{eq:delay}
\end{equation}
Equation~\eqref{eq:delay} holds under single-mode infrared control, which the independent phase-delay calculation tests directly.
The directly computed phase delay gives $p_{\rm IR}=0.994\pm0.002$ for $\qr\geq30$ (Fig.~\ref{fig:topology}).
Its density convergence, together with the full complex-spectrum agreement, identifies the DBI infrared response with nonlinear relaxation.
No symmetry used here enforces $p_{\rm IR}=1$ exactly.
The cancellation of $1-r^4$ in the linear delay integral at $z=2$ and the
dense-limit scaling of the DBI equations single out the same marginal radial
problem, which helps explain why a simple one-rate description becomes so
accurate.  They do not, however, constitute an analytic proof that the
constitutive residue vanishes in the cubic problem.  The value near unity is
therefore a dynamical result of the nonlinear radial evolution; the
nonmarginal comparison below shows explicitly that the DBI square root and
higher-form symmetry alone do not fix it.
Across the DBI family the topology weight remains dynamical:
for nonmarginal $z=1$ and $1.5$ backgrounds, the same low-frequency form fits with mixed weights $p\simeq0.65$ and $0.82$--$0.85$, respectively, as detailed below.

\label{sec:dbi-test}
We compare the numerical DBI response with both limiting topologies in
Eq.~\eqref{eq:pform}, fixing $\Gamma_J=1/\tau_J$ from the exact linear result.
For complex spectra $S_{\rm num}(\omega_n)$ and $S_{\rm EFT}(\omega_n)$ sampled on the numerical grid, the relative error is
\begin{equation}
 \epsilon_2=
 \left[
 \frac{\sum_{\omega_n\leq\Gamma_J}
 |S_{\rm num}(\omega_n)-S_{\rm EFT}(\omega_n)|^2}
 {\sum_{\omega_n\leq\Gamma_J}|S_{\rm num}(\omega_n)|^2}
 \right]^{1/2}.
\label{eq:L2error}
\end{equation}
The one-parameter finite-window fit minimizes the same norm over $\omega\leq\Gamma_J$ with only $p$ varied; varying the upper edge between $0.6\Gamma_J$ and $1.2\Gamma_J$ shifts $p_{\rm fit}$ by $0.014$, $0.0067$, and $0.0041$ for $\qr=10$, $30$, and $100$, respectively.
\begin{table}[t]
\caption{Pole-topology comparison.  The mixed fit varies only $p$; both pure predictions are parameter free after the linear response is known.}
\centering
\begin{tabular}{ccccc}
\toprule
$\qr$ & $p_{\rm fit}$ & mixed error & relaxation error & constitutive error\\
\midrule
3   & 1.198 & 0.109 & 0.148 & 0.619\\
10  & 0.997 & 0.038 & 0.038 & 0.558\\
30  & 0.973 & 0.022 & 0.028 & 0.588\\
100 & 0.975 & 0.015 & 0.022 & 0.632\\
\bottomrule
\end{tabular}

\end{table}
The fitted finite-window value approaches the relaxation class but retains
$O(\omega/\Lambda_{\rm fast})$ corrections.
For $\qr=3$, $p_{\rm fit}=1.198$ is physically a signed-weight decomposition:
$1-p_{\rm fit}=-0.198$ represents a constitutive contribution that partially cancels the relaxation contribution.
The independent low-density infrared trend also has $p_{\rm IR}>1$, while finite-window corrections affect the precise fitted magnitude.
Those corrections are diagnosed by the displacement of the finite-window values from the infrared trend.
The independent low-frequency estimator in Eq.~\eqref{eq:pir}, with the common factor $\rh^2$ cancelling in the ratio, is more direct:
for $\qr\geq30$ it gives $p_{\rm IR}=0.994\pm0.002$.
The four factors in the limiting response have a transparent origin:
three propagators dress the incoming fields and one dresses the emitted harmonic.
The same product structure occurs in uniform nonlinear electron hydrodynamics, where the slow variable is ordinarily momentum \cite{Sun2018}.
In the probe-brane metal it instead emerges from the approximately conserved electric-current/higher-form sector \cite{Chen2017,Davison2023}.

\paragraph{Nonmarginal DBI backgrounds.}\label{sec:nonmarginal}
To test whether $p\simeq1$ is imposed by the DBI square root itself, we repeat the calculation for $z=1$ and $z=1.5$.
Unlike at $z=2$, the linear rate is not known analytically and is obtained from a Drude-plus-incoherent fit to the linear conductivity only.
The cubic data then determine the single topology weight $p$ over $\omega\leq\Gamma_J$.
\begin{table}[t]
\caption{Mixed-topology test away from $z=2$.  The mixed error uses one cubic parameter $p$; $\Gamma_J$ is fixed by linear response.}
\centering
\begin{tabular}{cccccc}
\toprule
$z$ & $\qr$ & $p_{\rm IR}$ & $p_{\rm fit}$ & mixed error & constitutive error\\
\midrule
1.0 & 30  & 0.676 & 0.661 & 0.0011 & 0.316\\
1.0 & 100 & 0.650 & 0.646 & 0.0003 & 0.303\\
1.5 & 30  & 0.816 & 0.820 & 0.0025 & 0.596\\
1.5 & 100 & 0.835 & 0.854 & 0.0026 & 0.380\\
\bottomrule
\end{tabular}

\end{table}
The nonmarginal models realize genuine mixtures of constitutive and relaxation nonlinearities.
This cross-background result supports the two-topology EFT and shows that the nearly pure relaxation limit of the marginal $z=2$ metal is selected dynamically within the DBI family.
Likewise, the $T^{3/2}$ nonlinear field scale is a separate fingerprint of the $z=2$ DBI scaling structure rather than a consequence of the relaxation topology.
Field scaling and pole topology are therefore logically independent and experimentally complementary diagnostics.

\section{Nonlinear dc response and scaling}
\label{sec:scaling}
For $z=2$ and $\qr\gg1$,
\begin{equation}
 \sigma_{\rm dc}\simeq\frac{\rho}{\pi T},
\qquad
 \sthree_{\rm dc}\simeq-\frac{\rho}{2(\pi T)^4}.
\label{eq:Tscaling}
\end{equation}
The corresponding field at which cubic and linear currents compete is
\begin{equation}
 E_{\rm nl}\equiv
\sqrt{\left|\frac{\sigma_{\rm dc}}{\sthree_{\rm dc}}\right|}
=\sqrt2\,(\pi T)^{3/2}.
\label{eq:Enl}
\end{equation}
At finite density the exact result is
\begin{equation}
 E_{\rm nl}(T,\rho)=\sqrt2\,(\pi T)^{3/2}
 \sqrt{1+\left(\frac{\pi T\mathcal N}{\rho}\right)^2},
\label{eq:EnlFinite}
\end{equation}
so Eq.~\eqref{eq:Enl} is approached as
$\rho/(\pi T\mathcal N)\to\infty$.
Operationally, $E_{\rm nl}$ is obtained by equating the measured linear and cubic current magnitudes.

\begin{figure}[t]
\includegraphics[width=\linewidth]{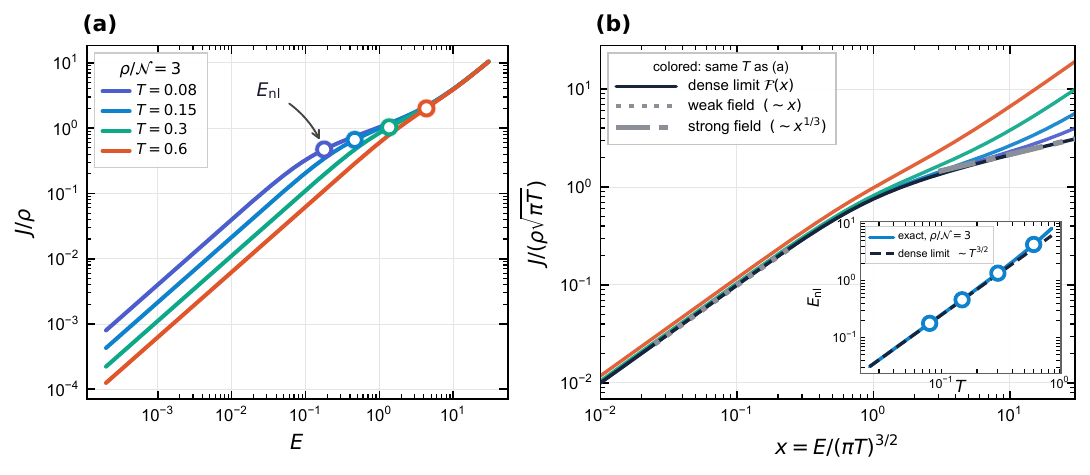}
\caption{\label{fig:dcscaling}
Nonlinear dc scaling of the $d=z=2$ DBI metal at fixed $\rho/\mathcal N=3$.
(a) Exact finite-density currents in the original variables; the four open circles mark $E_{\rm nl}(T)$ for their respective temperatures, where the linear and cubic current magnitudes are equal. The arrow identifies one representative marker.
(b) The same curves, expressed in the variables
$x=E/(\pi T)^{3/2}$ and
$J/(\rho\sqrt{\pi T})$, flow toward the parameter-free dense-limit function
$\mathcal F(x)=xy^2$, where $x^2y^6+y^4=1$.
Gray reference lines show its weak- and strong-field limits,
$x-x^3/2$ and $x^{1/3}$.
Each is shown over one decade inside its controlled regime:
$0.025\leq x\leq0.25$ and $3\leq x\leq30$, respectively.
At the largest scaled fields, the explicit $\widehat\rho^{-2}$ pair-production term produces the controlled finite-density departure from the dense curve.
Inset: the blue solid line is the exact finite-density $E_{\rm nl}(T,\rho)$ at the same fixed density, the blue open circles mark the four temperatures sampled in (a), and the black dashed line is the asymptotic dense-limit law $\sqrt{2}(\pi T)^{3/2}$.}
\end{figure}

The full nonlinear solution is organized by the dimensionless variables
\begin{equation}
 x=E\rh^3=\frac{E}{(\pi T)^{3/2}},
 \qquad y=\frac{r_\ast}{\rh},
 \qquad \widehat\rho=\frac{\rho\rh^2}{\mathcal N}
 =\frac{\rho}{\pi T\mathcal N}.
\end{equation}
In these variables the exact current is
\begin{equation}
 \frac{J}{\rho\sqrt{\pi T}}
 =x\sqrt{y^4+\widehat\rho^{-2}},
\qquad
 x^2y^6+y^4=1,
\label{eq:finitecollapse}
\end{equation}
Figure~\ref{fig:dcscaling} shows the full finite-density flow toward
the density-dominated function
\begin{equation}
 \mathcal F(x)=xy^2,
 \qquad x^2y^6+y^4=1,
 \label{eq:densefunction}
\end{equation}
not only its cubic expansion. Its controlled limits are
\begin{equation}
 \mathcal F(x)=x-\frac{x^3}{2}+O(x^5),
 \qquad
 \mathcal F(x)=x^{1/3}+O(x^{-1}),
 \label{eq:dcasymptotics}
\end{equation}
for $x\ll1$ and $x\gg1$, respectively.
The residual high-field separation is fixed by the
$\widehat\rho^{-2}$ pair-production term in Eq.~\eqref{eq:finitecollapse},
rather than by an additional fitted scale.
The nonlinear hydrodynamic basin therefore shrinks as $T^{3/2}$ even though its scaled shape becomes density independent.

As an independent, secondary diagnostic, the holographic crossover field scales as $T^{3/2}$, whereas the temperature-dependent YSYK current has an $E/T$ scaling regime in the field convention of Ref.~\cite{Kryhin2025}.
The holographic infrared spectrum also contains one rate $\Gamma_J$ and flows to the outgoing-pole topology.
The YSYK response instead resolves energy and quadrupolar deformation rates through products of distinct relaxation factors.
In the minimal electrothermal reduction, the pole locations are fixed by the separately measurable ratio $\Gamma_E/\Gamma_M$ and contain pair-frequency poles absent from the single-current DBI response.
These diagnostics are complementary: the field scaling probes the nonlinear scaling structure, whereas pole topology tests how the slow current becomes nonlinear.
The same field exponent need not imply the same pole topology, or conversely.

For a direct scaling test, at each temperature the measured current should first be divided by
$\rho\sqrt{\pi T}$ and the field by $(\pi T)^{3/2}$.
In the density-dominated regime, all curves should approach the same
$\mathcal F(x)$ near the onset of nonlinearity.
The field $E_{\rm nl}$ can be extracted independently by fitting the weak-field current to
$J=\sigma_{\rm dc}E+\sigma_{\rm dc}^{(3)}E^3+\cdots$ and forming
$\sqrt{|\sigma_{\rm dc}/\sigma_{\rm dc}^{(3)}|}$.
A systematic upward deviation from the $T^{3/2}$ law at higher temperature is the finite-density correction in Eq.~\eqref{eq:EnlFinite}, whereas an $E/T$ collapse would indicate a different strange-metal mechanism.
For Fig.~\ref{fig:dcscaling} we set $\rho/\mathcal N=3$ and use $T=0.08$, $0.15$, $0.30$, and $0.60$, corresponding to $\widehat\rho=11.94$, $6.37$, $3.18$, and $1.59$. The curves are direct evaluations of Eqs.~\eqref{eq:finitecollapse} and \eqref{eq:densefunction}; no fit parameters enter.

\section{Phenomenological implications}
\label{sec:phenomenology}
The holographic and material applications play distinct roles.  The former
provides a controlled realization of one pole topology, whereas the latter
illustrates how the effective theory can discriminate between candidate slow
sectors using independently measured rates.  No microscopic identification
of LSCO with the probe-brane theory is assumed.  Phase-sensitive nonlinear
THz methods are established in superconducting and correlated quantum
materials \cite{Matsunaga2014,ShimanoTsuji2020,Chu2023,Barbalas2025,Liu2025},
while a large normal-state third-order THz susceptibility and distinct Drude
and energy-relaxation rates have been reported in overdoped LSCO
\cite{Chaudhuri2026}.
If the pump--probe energy rate is the scalar channel entering the cubic feedback vertex, the reported interval $g=\Gamma_E/\Gamma_M=1/40$--$1/10$ gives $x_\times=0.0496$--$0.0969$ and $\Gamma_M\tau_{\rm THG}=25$--$85$, sharply separated from the one-current values $0.2918$ and $6$.

\begin{figure}[t]
\includegraphics[width=\linewidth]{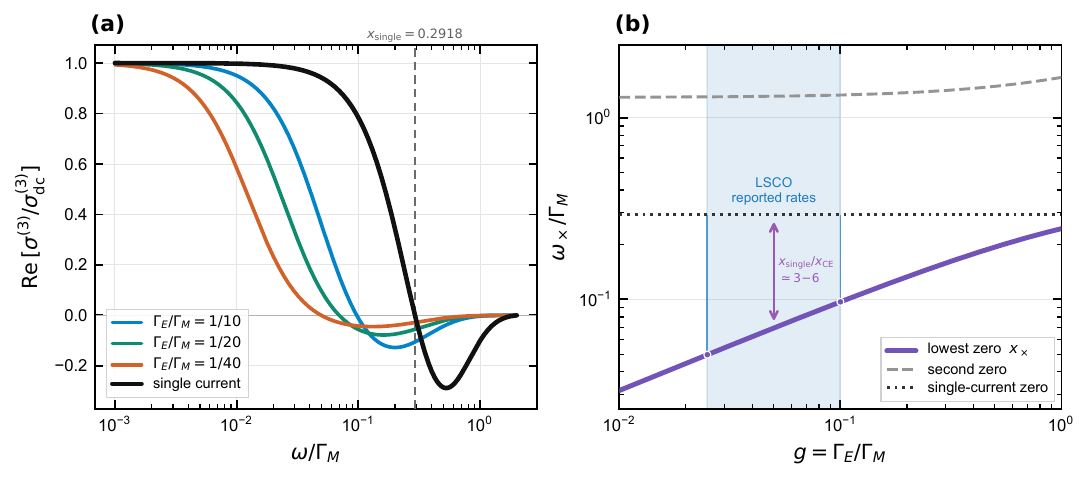}
\caption{\label{fig:twomode}
Two-mode pole topology constrained by independently determined rates.
(a) Normalized dissipative THG for a single relaxational current (black) and for a current coupled to a scalar energy mode at the indicated $g=\Gamma_E/\Gamma_M$.
The vertical dashed line marks the single-current zero.
(b) The two positive THG zeros of the current--energy topology.
The shaded band denotes the rate separation $g=1/40$--$1/10$ reported for normal-state LSCO \cite{Chaudhuri2026}. Within the minimal scalar reduction this interval places the lower zero at $\omega_\times/\Gamma_M=0.0496$--$0.0969$.
The finite endpoint lines connect the lower-zero prediction at each rate boundary to the single-current zero; the small purple circles mark their intersections with the lower-zero curve. The purple arrow shows the corresponding ratio $x_{\rm single}/x_{\rm CE}\simeq3$--$6$ between the one-current and current--energy zeros. No nonlinear amplitude is fitted.}
\end{figure}

Measurements above $T_c$, or in extremely overdoped films with very low $T_c$, suppress Higgs, Josephson-plasma, and superconducting fluctuation backgrounds.
Pseudogap, stripe, and charge-order fluctuations in La-based cuprates may instead add poles or smooth nonlinear backgrounds \cite{Missiaen2025}.
First determine $\Gamma_M(T)$ from the complex linear conductivity and
$\Gamma_E(T)$ from the pump--probe decay, then sweep the drive frequency and
record both THG quadratures.
For a Planckian current rate $\Gamma_M=\alpha k_BT/\hbar$ \cite{Legros2019,HartnollMackenzie2022}, the single-current prediction is
\begin{equation}
 \nu_\times=\frac{\omega_\times}{2\pi}
 =0.2918\,\alpha\frac{k_BT}{h}
 \simeq 6.08\,\alpha\,T[{\rm K}]\ {\rm GHz}.
\end{equation}
Thus $30$--$100$ K corresponds to roughly $0.18$--$0.61\,\alpha$ THz, within modern frequency-agile THz sources.
The minimal electrothermal prediction instead lies at
$0.0496$--$0.0969$ times $\Gamma_M$ for the reported LSCO rate ratios.
Experimental uncertainty can be propagated without introducing a nonlinear
fit.  Writing the predicted absolute zero as
$\omega_\times=\Gamma_M x_\times(g)$ with
$g=\Gamma_E/\Gamma_M$, independent rate uncertainties give
\begin{equation}
 (\delta\omega_\times)^2
 =\left[x_\times-gx_\times'(g)\right]^2(\delta\Gamma_M)^2
  +\left[x_\times'(g)\right]^2(\delta\Gamma_E)^2 .
\label{eq:zero_uncertainty}
\end{equation}
For $g\ll1$, $\omega_\times\simeq\sqrt{\Gamma_E\Gamma_M/10}$, so
\begin{equation}
 \frac{\delta\omega_\times}{\omega_\times}
 \simeq\frac12\sqrt{
 \left(\frac{\delta\Gamma_E}{\Gamma_E}\right)^2+
 \left(\frac{\delta\Gamma_M}{\Gamma_M}\right)^2},
\label{eq:zero_uncertainty_smallg}
\end{equation}
with the covariance term restored when the two rates are extracted from
correlated data.
Observing either interval would support the corresponding minimal slow-sector description;
additional phase structure can be analyzed with the general multimode theory.

The disorder threshold depends on the residues and relaxation scales of the
added modes.
For a smooth additive background
$\delta\sigma^{(3)}(\omega)$, however, the leading displacement of a simple
in-phase zero is
\begin{equation}
 \delta\omega_\times
 =-\frac{\operatorname{Re}\delta\sigma^{(3)}(\omega_\times)}
 {\partial_\omega\operatorname{Re}\sigma^{(3)}_0(\omega_\times)} .
\label{eq:zero_shift}
\end{equation}
The topology marker is experimentally resolved when this shift, combined
with Eq.~\eqref{eq:zero_uncertainty}, is smaller than the separation between
the candidate zero intervals.  A narrow disorder or phonon pole is not a
smooth background and must instead be included as an additional mode in
$\bm{\Gamma}$.

\section{Discussion}
\label{sec:discussion}

Phase-resolved third-order spectroscopy can diagnose the slow sector of a strange metal through pole topology in the leading low-frequency regime where linear transport alone is ambiguous.
Approximate conservation produces a slow current, while the third-order pole topology reveals whether nonlinearity resides in its equation of state or in the mechanism that breaks the conservation law.
Higher-form symmetry supplies one realization of the slow current but does not fix its cubic coefficients.
This matters for strange metals because $T$-linear resistivity alone does not identify the slow sector, whereas nonlinear pole topology tests its dynamics.

The DBI strange metal provides a nonperturbative example, within the classical large-$N$ probe sector, in which higher-form symmetry explains the slow current while the DBI dynamics dynamically selects nonlinear relaxation; finite-$N$ quantum fluctuations are not included.
The same leading-pole logic applies to nearly conserved currents, phase-relaxed superfluids, and other slow-mode conductors.
The result is therefore not a statement about DBI alone: the holographic calculation supplies a controlled microscopic realization of one topology, while the effective theory organizes the broader class of possible slow-sector nonlinearities.

Backreaction is a substantive extension rather than a harmless technical
correction.  At finite flavour-to-colour ratio the brane stress tensor mixes
the probe current with energy and momentum, while a lattice or axion sector
adds explicit momentum relaxation; backreacted DBI--axion models already show
that even the dc conductivity matrix is richer than in the probe limit
\cite{CremoniniHooverLi2017}.  The pole-topology language remains applicable
through the relaxation matrix of section~\ref{sec:eft}, but the isolated
one-current zero at $0.2918\Gamma_J$ is not protected.  It is perturbatively
stable only when the probe-current eigenmode remains parametrically separated
from the momentum and energy rates and their mixing residues are small.  If
momentum becomes comparably slow, additional poles and shifted or absent
zeros are expected.  Establishing the topology in that regime requires a
fully backreacted cubic fluctuation calculation.

A direct experimental protocol is: (i) extract $\Gamma_M$ from the complex linear THz conductivity; (ii) determine $\Gamma_E$ from pump--probe decay; and (iii) sweep phase-resolved THG to locate the zero of its in-phase component.
Repeating the third step for fields along the crystal axes and diagonals, and measuring the transverse cubic current at intermediate angles, separates scalar heating from $B_{1g}$ and $B_{2g}$ deformation channels \cite{Kryhin2025,Watanabe2025}.
Extra poles from disorder, phonons, competing-order fluctuations, or additional slow modes diagnose departure from the one-current form in Eq.~\eqref{eq:pform}.

$La_{2-x}Sr_xCuO_4$ films provide a concrete platform: their strange-metal transport and Planckian phenomenology are well established \cite{Cooper2009,Legros2019,HartnollMackenzie2022}, phase-sensitive nonlinear THz methods are established in superconducting and correlated quantum materials \cite{Matsunaga2014,ShimanoTsuji2020,Chu2023,Barbalas2025,Liu2025}, and the LSCO normal state has a large third-order THz response with independently measured $\Gamma_M$ and $\Gamma_E$ \cite{Chaudhuri2026}.
If those rates enter the minimal scalar current--energy feedback channel, they place the current--energy zero at $0.050$--$0.097\Gamma_M$, well below the single-current value $0.2918\Gamma_M$.
Mixed-frequency 2D spectroscopy is still more decisive because Eqs.~\eqref{eq:ceu}--\eqref{eq:cee} predict ridges controlled by $\omega_i+\omega_j$, whereas the DBI relaxation topology contains only incoming and total outgoing current poles \cite{Wan2019,Salvador2024}.
Measurements above $T_c$ or in low-$T_c$ overdoped films reduce superconducting collective-mode backgrounds \cite{Chu2023,Salvador2024}.
Pseudogap, stripe, or charge-order fluctuations may add poles or smooth nonlinear backgrounds; such structures signal physics beyond the single-current form and can be incorporated as additional slow modes \cite{Delacretaz2017PRB,Missiaen2025}.

\section*{Data and code availability}
No public data or code release accompanies this article. The results reported here are derived from the equations and numerical procedures specified in the text.

\bibliographystyle{JHEP}
\bibliography{refs}
\end{document}